\documentclass{IEEEcsmag}

\usepackage[colorlinks,urlcolor=blue,linkcolor=blue,citecolor=blue]{hyperref}
\usepackage{hyperref}
\expandafter\def\expandafter\UrlBreaks\expandafter{\UrlBreaks\do\/\do\*\do\-\do\~\do\'\do\"\do\-}
\usepackage{upmath}

\usepackage[utf8]{inputenc}
\usepackage[dvipsnames]{xcolor}

\jvol{XX}
\jnum{XX}
\paper{8}
\jmonth{Month}
\jname{IEEE Security \& Privacy}
\jtitle{Execution and assessment of agentic influence operations in simulated social networks}
\pubyear{2026}

\begin{document}


\sptitle{Theme Article: Autonomous AI Agents in Computer Security}

\title{Execution and assessment of agentic influence operations in simulated social networks}

\author{Alejandro Buitrago L\'opez}
\affil{Dept. of Information and Communications Engineering, University of Murcia, 30100, Spain}

\author{David~Montoro~Aguilera}
\affil{Dept. of Information and Communications Engineering, University of Murcia, 30100, Spain}

\author{Javier Pastor-Galindo}
\affil{Dept. of Information and Communications Engineering, University of Murcia, 30100, Spain}

\author{Jos\'e~A.~Ruip\'erez-Valiente}
\affil{Dept. of Information and Communications Engineering, University of Murcia, 30100, Spain}

\markboth{Autonomous AI Agents in Computer Security}{Autonomous AI Agents in Computer Security}

\begin{abstract}This article evaluates AI-enabled influence operations in synthetic social networks through controlled simulations of narrative release, amplification, and counter-messaging. We measure exposure and belief change in agentic audiences, showing that amplification maximizes reach, counter-messaging shifts opinions most, and narrative release requires larger attacker footprints.


\end{abstract}

\maketitle

\chapteri{O}nline social networks (OSNs) are now part of the security landscape. Beyond enabling communication and news discovery, they provide an attack surface for influence operations (IOs). Adversaries can seed narratives, coordinate engagement, and exploit platform amplification to shape exposure and downstream attitudes \cite{vosoughi2018}. These campaigns increasingly appear within broader hybrid operations, where informational effects are combined with reusable tactics, techniques, and procedures (TTPs) and supported by shared infrastructure \cite{larson2009}. Recent large language models (LLMs) further expand this capability by lowering the cost of generating persuasive, audience-tailored content and rapidly varying narrative framing \cite{bontridder2021ai,llmsecurity2024,llmcyber2024}.

In this context, rigorous evaluation of offensive manipulation remains difficult on real platforms. Most evidence is observational, making it difficult to evaluate what would have happened under different conditions: campaigns cannot be rerun with different attacker budgets, coordination strategies, or interventions, and the mechanisms shaping exposure remain only partially observable \cite{lazer2020css,guess2019less}. Simulation with AI agents helps address part of this limitation \cite{pastorgalindo2024llmframework}. Nevertheless, prior work emphasizes scale or behavioral realism of social phenomena \cite{oasis2024,beskow2019abm,botsim2024}, rather than a controlled, security-oriented assessment of adversarial strategies, red teaming, and their effects. This matters because it allows different influence strategies to be compared under the same conditions on both reach and persuasion, while also enabling a systematic evaluation of whether defensive interventions reduce adversarial exposure, belief change, or both.

In this study, we deploy offensive AI agents in simulated social networks with three different influence mechanisms: \textit{Narrative Release}, \textit{Narrative Support}, and \textit{Counter-Narrative Reaction} \cite{pastorgalindo2025iosn}. To preserve interpretability, agent behavior is programmed with deterministic state machines and a LLM is used for the creation of organic content. Then, we evaluate the reach of narratives on the social network and the belief shift of the synthetic audiences. Across experiments that vary the proportion of red accounts, we quantify operation effectiveness through exposure (IO reach) and belief change, measured with Jensen--Shannon divergence.

\section{SIMULATED SOCIAL NETWORKS}

\begin{figure}[h!]
    \centering
    \includegraphics[width=\columnwidth]{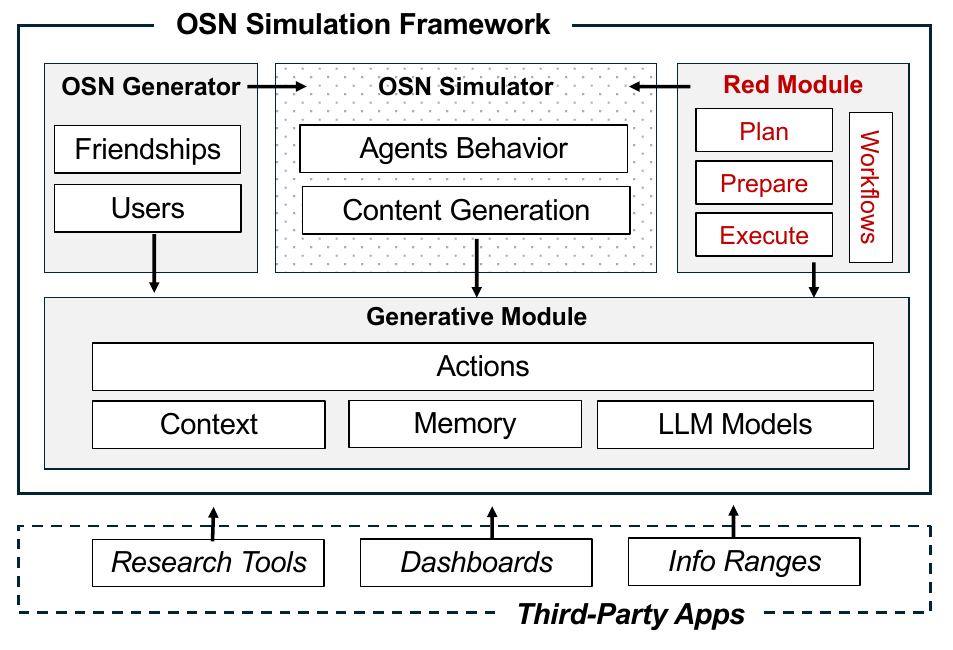}
    \caption{Overview of the proposed framework for simulating OSNs and IOs \cite{lopez2025agentbasedsimulationonlinesocial}}
    \label{fig:framework}
\end{figure}

To support the controlled evaluation of IOs, we rely on an agent-based OSN simulation framework described in a previous work \cite{lopez2025agentbasedsimulationonlinesocial}. The simulator separates network generation, agent behavior, content generation, and adversarial execution, as shown in Figure~\ref{fig:framework}.

Given a new simulation and before any content is produced, the \textit{OSN Generator} instantiates a directed follower graph with cyberpersonas that are demographically representative (name, age, gender, traits, interests, and occupation) and realistic connectivity patterns \cite{11441429}. Then, the \textit{OSN Simulator} is launched through typical platform actions such as reading, posting, replying, sharing, and liking. An agent is exposed to posts and replies from accounts the agent follows in the previously generated social graph. The simulated user behavior is coded with an automaton that determines when and how these actions occur, where the probabilities are based on demographic attributes of the synthetic users \cite{lopez2025agentbasedsimulationonlinesocial}. They also have a short-term and long-term memory module for coherence. Moreover, the \textit{Generative Module} is configured with structured prompting and the Gemini API. It is used by agents only for creating organic content when a reply or writing action is needed.
      
Adversarial behavior is introduced via the \textit{Red Module}, which extends the framework without altering the basic mechanics of the simulated platform. It is programmed with three workflows that manage the actions of a subset of accounts of the simulation for offensive purposes (red agents). Red accounts still operate through the simulator algorithm, but their individual decisions are guided by one of the workflow's logic paths, which differs from normal functioning. For content generation, the \textit{Generative Module} uses the \textit{gemini-2.5-flash} model with a custom prompting strategy. 

The framework captures the full stream of simulation events, including users, links, interactions, and timestamps, to assess reach and belief changes. In the next section, we describe in detail how artificial influence operations are launched.

\section{RED TEAMING OF INFLUENCE OPERATIONS}
\label{sec:red}

\begin{figure*}[t]
    \centering
    \includegraphics[width=0.8\textwidth]{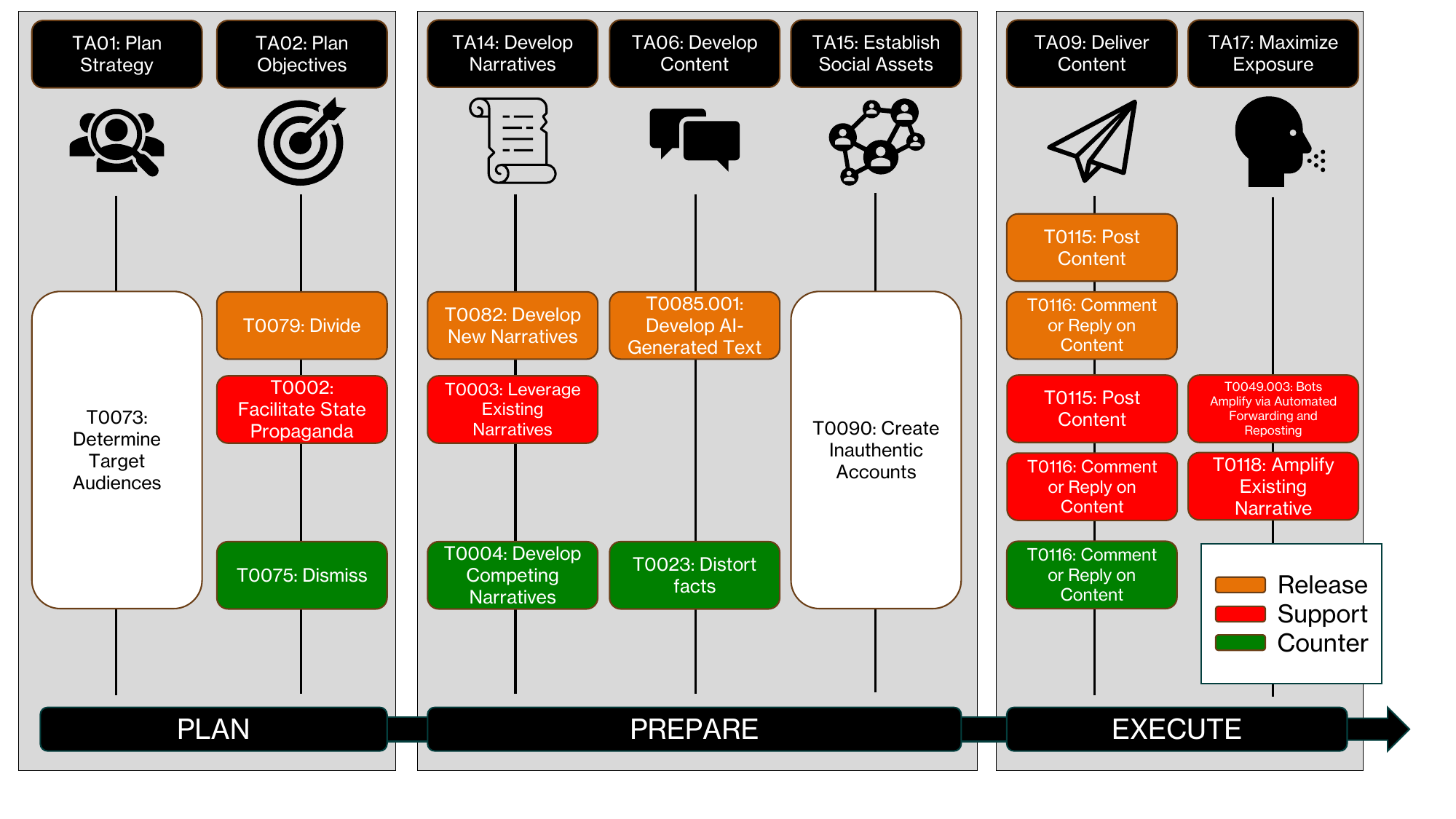}
    \caption{DISARM tactics (TA) and techniques (T) of the implemented workflows across the \textit{Plan}, \textit{Prepare}, and \textit{Execute} phases.}
    \label{fig:workflow_ttps}
\end{figure*}

Each operation is launched within the simulator of synthetic social networks through an explicit campaign specification. Before execution, the red operator defines:

\begin{itemize}
    \item the target simulated social network (previously created in the simulator)
    \item the discussion topic (influencing general theme)

    \item the workflow to deploy (namely, \textit{Narrative Release}, \textit{Narrative Support} or \textit{Counter-Narrative Reaction})
\end{itemize}

This parameterization allows varying the presence of attackers and the campaign design without changing the underlying simulator. To execute coordinated campaigns, we model step-by-step explicit workflows executed by malicious inserted users. They will be deployed against synthetic audiences in realistic social graph and evaluated afterwards.

\subsection{Workflows of influence}

In this study, three workflows are designed: \textit{Narrative Release}, \textit{Narrative Support} and \textit{Counter-Narrative Reaction} \cite{pastorgalindo2025iosn}. Each workflow is a set of ordered tactics, techniques, and procedures (TTPs) based on the DISARM Red Framework \cite{terp2022disarm}, since it provides a standardized and modular description of online operations patterns \cite{pastorgalindo2025iosn}.

Figure~\ref{fig:workflow_ttps} summarizes these workflow compositions across the DISARM phases of plan, prepare and execute. 

All three workflows share two common techniques (white boxes of the image): \textit{T0073: Determine Target Audiences}, used in the plan phase to define the intended audience of the operation that match the demographic attributes of accounts; and \textit{T0090: Create Inauthentic Accounts}, configured in the prepare phase for the creation and insertion of red-controlled accounts in the target social network that will launch the actions of the execute phase. On top of this common basis, each workflow operationalizes a different set of workflow-specific TTPs.

\subsubsection{\textbf{\textcolor{orange}{Narrative Release}}}

\textit{Narrative Release} introduces a new narrative into the network. As shown in Figure~\ref{fig:workflow_ttps}, it follows an ordered sequence of DISARM techniques across the \textit{Plan}, \textit{Prepare}, and \textit{Execute} phases.

In \textit{Plan}, the operator defines the topic, target audience, and objective. The audience is specified through \textit{T0073: Determine Target Audiences}, and the main objective is \textit{T0079: Divide}.

In \textit{Prepare}, the narrative is created through \textit{T0082: Develop New Narratives}. Example content may also be generated through \textit{T0085.001: Develop AI-Generated Text}. Red-controlled accounts are then created through \textit{T0090: Create Inauthentic Accounts}. At this stage, the user completes the main configuration: topic, target audience, objective, narrative, and number of red agents.

In \textit{Execute}, the workflow is deployed through \textit{T0115: Post Content} and \textit{T0116: Comment or Reply on Content}. Once the simulation starts, red agents spread the selected narrative through ordinary posts and replies.

This workflow follows the standard sequential flow of the simulator. Red agents do not identify narratives at runtime, analyse other agents' posts, or trigger reactive actions. They use the same actions as ordinary accounts, but their content is conditioned to promote the configured narrative. Runtime details are described later in the \textit{Runtime execution} subsection.

\subsubsection{\textcolor{red}{\textbf{Narrative Support}}}

\textit{Narrative Support} reinforces a narrative that is already present in the network. As shown in Figure~\ref{fig:workflow_ttps}, it is also organized across the \textit{Plan}, \textit{Prepare}, and \textit{Execute} phases.

In \textit{Plan}, the operator defines the topic, target audience, and objective. The audience is specified through \textit{T0073: Determine Target Audiences}, and the main objective is \textit{T0002: Facilitate State Propaganda}. Unlike \textit{Narrative Release}, the final target narrative is not fixed during configuration.

In \textit{Prepare}, the workflow is set to build on existing narratives through \textit{T0003: Leverage Existing Narratives}. Red-controlled accounts are created through \textit{T0090: Create Inauthentic Accounts}. At this stage, the operator defines the campaign settings, while the narrative to support will be selected later from the discussion.

In \textit{Execute}, the workflow uses \textit{T0115: Post Content}, \textit{T0116: Comment or Reply on Content}, \textit{T0118: Amplify Existing Narrative}, and \textit{T0049.003: Bots Amplify via Automated Forwarding and Reposting}. Red agents first observe the discussion and identify the dominant narrative. They then support it by creating aligned posts and replies, and by boosting aligned benign content through likes and shares.

At the behavioural level, \textit{Narrative Support} follows an asynchronous strategy. Red agents begin in passive observation mode, identify the target narrative at runtime, and then amplify aligned content when it appears. What changes here is both the content they generate and their action policy. Runtime details are described later in the \textit{Runtime execution} subsection.

\subsubsection{\textbf{\textcolor{ForestGreen}{Counter-Narrative Reaction}}}

\textit{Counter-Narrative Reaction} seeks to undermine an existing narrative by promoting an opposing one. As shown in Figure~\ref{fig:workflow_ttps}, it also follows an ordered sequence across the \textit{Plan}, \textit{Prepare}, and \textit{Execute} phases.

In \textit{Plan}, the operator defines the topic, target audience, and objective. The audience is specified through \textit{T0073: Determine Target Audiences}, and the main objective is \textit{T0075: Dismiss}. As in \textit{Narrative Support}, the final target narrative is not fixed during configuration.

In \textit{Prepare}, the workflow sets up an alternative framing through \textit{T0004: Develop Competing Narratives}. Red-controlled accounts are created through \textit{T0090: Create Inauthentic Accounts}. The operator completes the campaign settings at this stage, while the final counter-narrative is completed once the target narrative is identified at runtime.

In \textit{Execute}, the workflow is deployed through \textit{T0023: Distort Facts} and \textit{T0116: Comment or Reply on Content}. Red agents first observe the discussion and identify the dominant narrative. That narrative becomes the one to counter. A competing narrative is then generated, and red agents reply only when a message is sufficiently aligned with the target narrative.

This workflow follows an asynchronous, reply-only strategy. Red agents begin in passive observation mode, identify the narrative to undermine, and then attach counter-messages to aligned posts. They do not aim to like, share, or amplify content. Runtime details are described later in the \textit{Runtime execution} subsection.

\subsection{Generation of influencing messages}

The LLM layer supports offensive content generation in the narrative and content-development techniques of the workflows, including \textit{T0082}, \textit{T0004}, and \textit{T0085.001}. Each writing or replying request from a red agent is built from a four-part prompt template:

\begin{itemize}
    \item \textit{System prompt}, which encodes the red agent demographic attributes (cyberpersona)
    \item \textit{Target prompt}, which specifies the intended target audience (\textit{T0073: Determine Target Audiences})
    \item \textit{Objective prompt}, which states the operational goal of the workflow (Plan)
    \item \textit{Task prompt}, which defines the topic and narrative to be promoted or opposed.
\end{itemize}

Together, these components enable the same workflow to generate posts and replies that remain consistent with each red agent's persona while staying aligned with the campaign objective. In the reactive workflows, the same LLM layer is also used for bounded analytic tasks, namely, runtime narrative identification, post-to-narrative alignment analysis, and, in the countering workflow, counter-narrative generation.

\subsection{Runtime execution}

\begin{figure}[h!]
    \centering
    \includegraphics[width=\columnwidth]{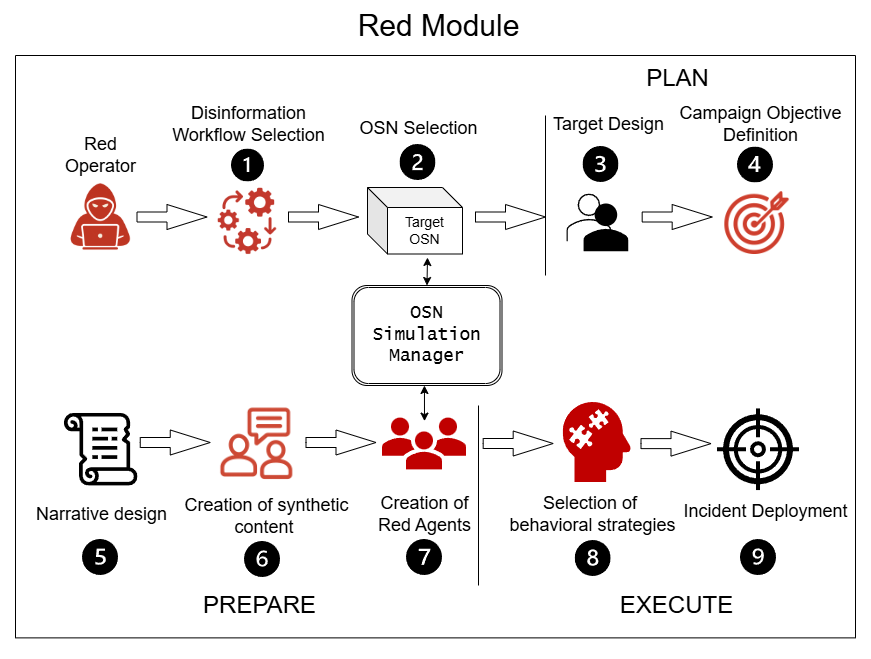}
    \caption{Workflow configuration and deployment through the integration of the Red Module and the OSN simulator. The process covers workflow selection, OSN selection, target design, objective definition, narrative and content preparation, red-agent creation, and final incident deployment.}
    \label{fig:red_agents_diagram}
\end{figure}

Figure~\ref{fig:red_agents_diagram} summarizes how an IO workflow is configured and deployed through the integration of the Red Module and the OSN simulator. The process follows the main DISARM phases used in this paper and is shared by the three workflows. In simple terms, the operator selects the workflow and target OSN, defines the target audience and objective, prepares the narrative and example content, creates the red agents, selects the behavioural strategy, and finally deploys the incident in the simulation.

The process starts with workflow selection and OSN selection. The operator chooses one of the three workflows and selects the target simulated social network, either by using an existing OSN or by creating a new one through the OSN simulator. In the \textit{Plan} phase, the operator then defines the topic, the target audience, and the workflow objective. The target audience is designed from the demographic attributes already available in the simulator, and this choice affects both the red agents created for the operation and the style of the LLM-generated content.

In the \textit{Prepare} phase, the operator designs the narrative to be used in the operation and may generate example social media content with the LLM. This content is not posted directly in the simulation, but serves as a preview of the style and framing that will later be used by the red agents. At this stage, the red-controlled accounts are also created and inserted into the target network. These accounts are built as ordinary social agents, but their prompting and execution policy are later adapted to the selected workflow.

The last step is the \textit{Execute} phase. Here, the configured incident is deployed and the selected behavioural strategy determines how red agents act once the simulation starts. This is where the main difference between proactive and reactive workflows appears.

In the proactive workflow, namely \textit{Narrative Release}, the target narrative is fixed during configuration. Once the simulation begins, red agents follow the standard sequential simulator flow and disseminate the configured narrative through ordinary posts and replies. They do not identify narratives at runtime, analyse other agents' posts, or wait for triggering events. In this case, what changes is mainly the content they generate. 

In the reactive workflows, namely \textit{Narrative Support} and \textit{Counter-Narrative Reaction}, the final target narrative is not fixed in advance. Instead, red agents begin in a passive observation stage. They first collect an initial sample of benign posts, identify the dominant narrative in the discussion, and then monitor new post and reply events. Each observed message is analysed with respect to the selected target narrative, and a workflow-specific reaction is triggered only when the alignment is high enough. In \textit{Narrative Support}, the reaction consists of amplification through aligned posts, supportive replies, likes, and shares. In \textit{Counter-Narrative Reaction}, the reaction is a reply-only counter-message built to undermine the original framing and promote a competing narrative. 

This execution process makes the role of both components explicit. The Red Module is responsible for configuring and deploying the operation, while the OSN simulator provides the environment in which red agents and standard agents interact. The proactive workflow uses a direct content-delivery strategy, whereas the reactive workflows add an extra layer of observation, narrative analysis, and event-driven reaction.

\section{IMPACT OF SIMULATED INFLUENCE OPERATIONS}
\label{sec:impact}

This section evaluates the three workflows under controlled conditions, focusing on their effects on audience reach and opinion change.

\subsection{Experimental setup}

All experiments use the same simulated environment, keeping conditions fixed so that differences in outcome can be linked to workflow design. The network size is set to 1{,}000 agents, following prior experimental testing that identified this value as a practical balance between social network complexity and acceptable runtime.

The proportion of red-controlled accounts varies across seven settings: 0\%, 4\%, 6\%, 8\%, 15\%, 30\%, and 45\%. This is the main experimental parameter, as it controls the density of the IO. The 0\% condition provides the baseline, while the remaining settings progressively increase attacker presence. Higher densities are included as controlled stress-test conditions. Each configuration is executed five times with different random seeds, a choice made to balance result consistency and runtime.

The three workflows are instantiated on different campaign-topic settings. \textit{Narrative Release} introduces a fixed immigration-related narrative. \textit{Narrative Support} amplifies a dominant narrative in discussion about China, Europe, and the United States. \textit{Counter-Narrative Reaction} targets runtime-identified narratives in discussion about the Russia--Ukraine war. For this reason, comparisons across workflows should be read as comparisons within the evaluated scenario set, not as topic-independent estimates of workflow superiority.

\subsection{Impact metrics}

Effectiveness is evaluated by separating visibility from persuasion. We propose two metrics.

\subsubsection{\textbf{IO reach}}

This metric measures the fraction of benign agents that consume adversarial content at least once. In practice, this is computed from simulator read events associated with messages authored by red-controlled accounts.

\subsubsection{\textbf{Belief change}}
This metric measures whether exposure translates into audience opinion updating after being reached by workflow messages.

The simulator does not assume a fixed rule for how opinions change. Instead, belief change is estimated after exposure by comparing belief vectors elicited before and during the simulation. This matters because seeing adversarial content is not the same as being persuaded by it.

Belief change is assessed only for benign agents reached by the operation, excluding red-controlled accounts. For each assessed agent, the IO target narrative is treated as the statement under evaluation. A prior belief vector \(P\) is elicited before message exposure, and four posterior vectors \((Q_1,Q_2,Q_3,Q_4)\) are elicited after cumulative exposure at 25\%, 50\%, 75\%, and 100\% of the simulation timeline. At each checkpoint, the elicitation prompt conditions the model on the agent's cyberpersona and the set of posts read up to that point.

Each belief vector is obtained by asking the model five repeated questions about the same statement: two Likert-style questions about whether the statement is true or false, and three binary yes/no or true/false questions. The model returns a probability distribution over answer options for each question, and these probabilities are converted into weighted averages to form a 5-dimensional belief vector. Likert responses are mapped to \([0, 0.25, 0.5, 0.75, 1]\), while binary responses are mapped to \([1, 0]\). Prior and posterior vectors are then normalized before Jensen--Shannon divergence is computed. The final belief-change score is reported as \(JSD(P,Q_4)\), while \(JSD(P,Q_t)\) is used to analyze intermediate change over time.

\subsection{\textcolor{orange}{Narrative Release}}

\begin{figure}[h!]
    \centering
    \includegraphics[width=\columnwidth]{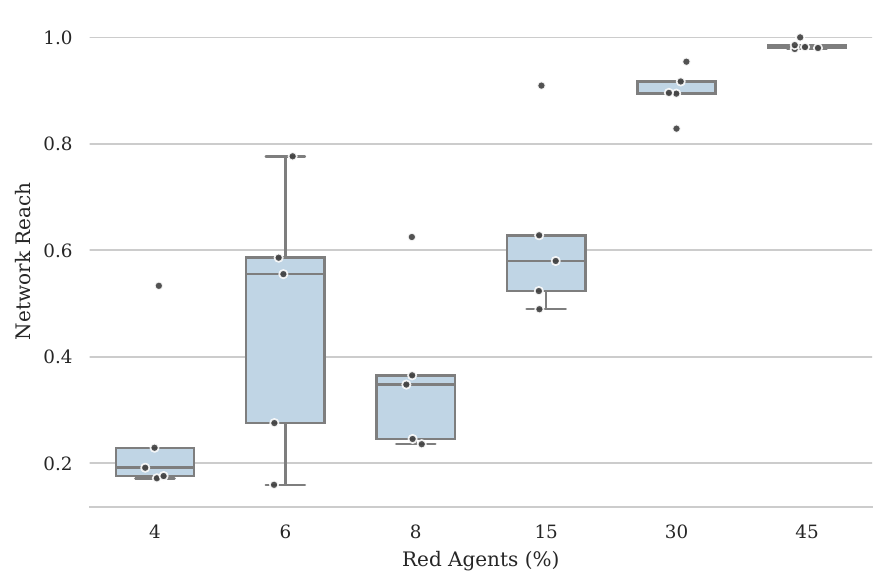}
    \caption{Narrative Release: Distribution of the network reached (IO reach) in five simulations per red agent configuration.}
    \label{fig:release_reach}
\end{figure}

This strategy introduces a narrative into the network. Figure~\ref{fig:release_reach} shows that reach is low and variable at small attacker footprints, but increases and becomes more stable as attacker presence grows. It exceeds 80\% at 30\% red-controlled accounts and is close to full-network exposure at 45\%.

These results show that proactive seeding can achieve broad visibility under the evaluated conditions, but only when the operation controls a sufficiently large share of accounts. In this sense, \textit{Narrative Release} behaves as a volume-dependent workflow.

\begin{figure}[h!]
    \centering
    \includegraphics[width=\columnwidth]{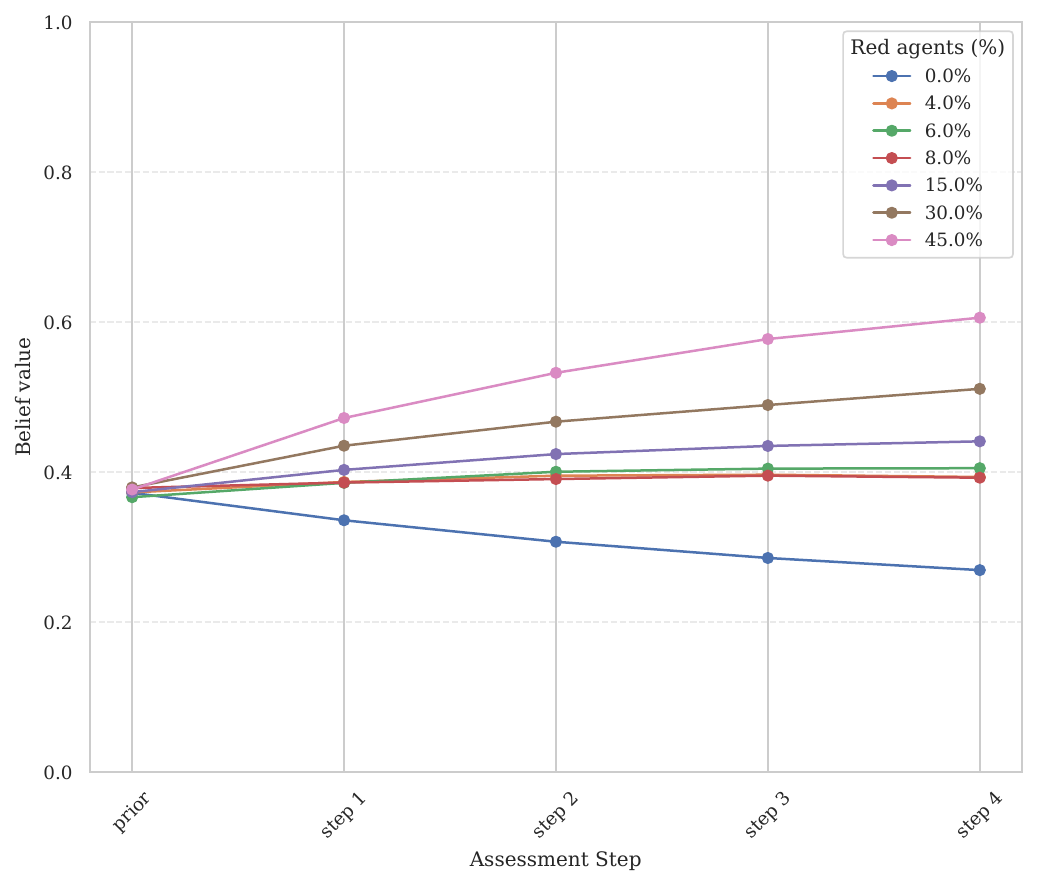}
    \caption{Narrative Release: Mean belief value evolution towards the target narrative.}
    \label{fig:release_belief}
\end{figure}

Figure~\ref{fig:release_belief} shows that baseline agreement with the injected narrative declines over time. Low-density release mainly slows this decline, whereas higher-density release increases agreement.

Taken together, these results suggest that \textit{Narrative Release} is a volume-driven mode. Under the tested conditions, it can move beliefs, but mainly when repeated exposure is sustained at scale.

\subsection{\textcolor{red}{Narrative Support}}

\begin{figure}[h!]
    \centering
    \includegraphics[width=\columnwidth]{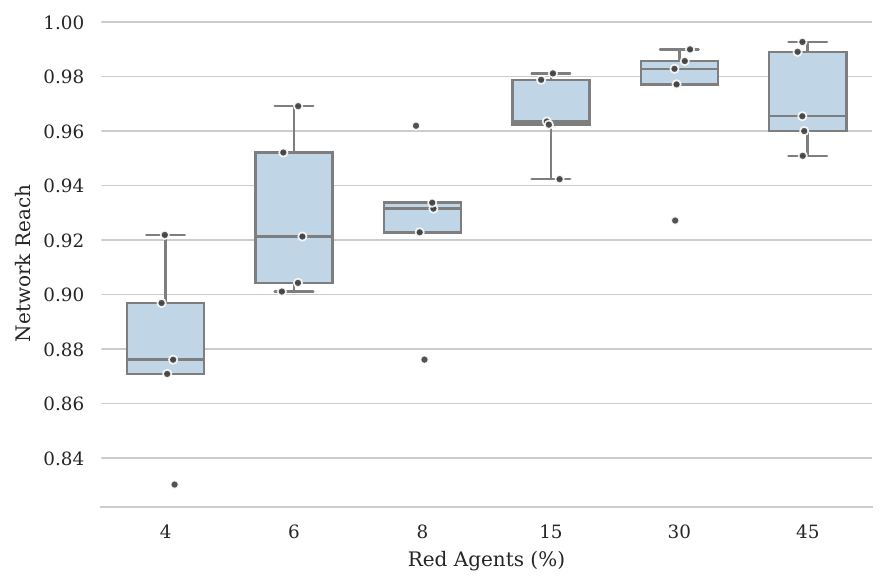}
    \caption{Narrative Support: Distribution of the IO reach (five simulations) per red agent configuration.}
    \label{fig:support_reach}
\end{figure}

\textit{Narrative Support} amplifies a narrative that is already prevalent among benign agents. Figure~\ref{fig:support_reach} shows a different exposure pattern from \textit{Narrative Release}: reach is high across all densities, typically above 0.80 and often close to full-network coverage. Rather than creating attention from scratch, controlled accounts attach themselves to already visible aligned content and amplify it through coordinated engagement.

As a result, even small attacker footprints can achieve broad exposure. In visibility terms, \textit{Narrative Support} is therefore more efficient than \textit{Narrative Release} at low densities.

\begin{figure}[h!]
    \centering
    \includegraphics[width=\columnwidth]{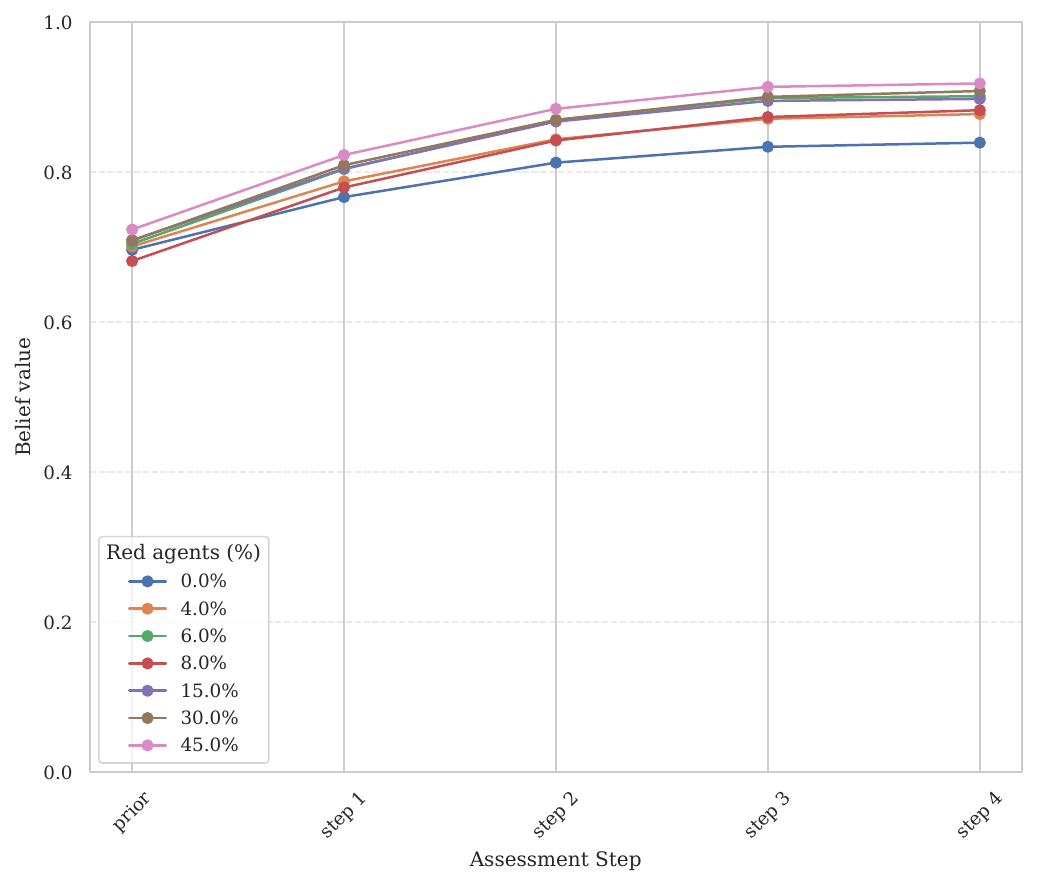}
    \caption{Narrative Support: Mean belief value evolution towards the target narrative.}
    \label{fig:support_belief}
\end{figure}

The belief results, however, are more limited. Figure~\ref{fig:support_belief} shows that agreement with the supported narrative rises even in the baseline, which suggests that the selected framing is already self-reinforcing under organic interaction. Additional adversarial amplification produces only modest gains beyond that trend.

Relative to \textit{Narrative Release}, this workflow is better at saturating exposure than at generating additional persuasion. Under the evaluated conditions, amplification is effective for visibility, but it adds limited marginal influence when the target narrative already benefits from organic reinforcement.

\subsection{\textcolor{ForestGreen}{Counter-Narrative Reaction}}

\begin{figure}[h!]
    \centering
    \includegraphics[width=\columnwidth]{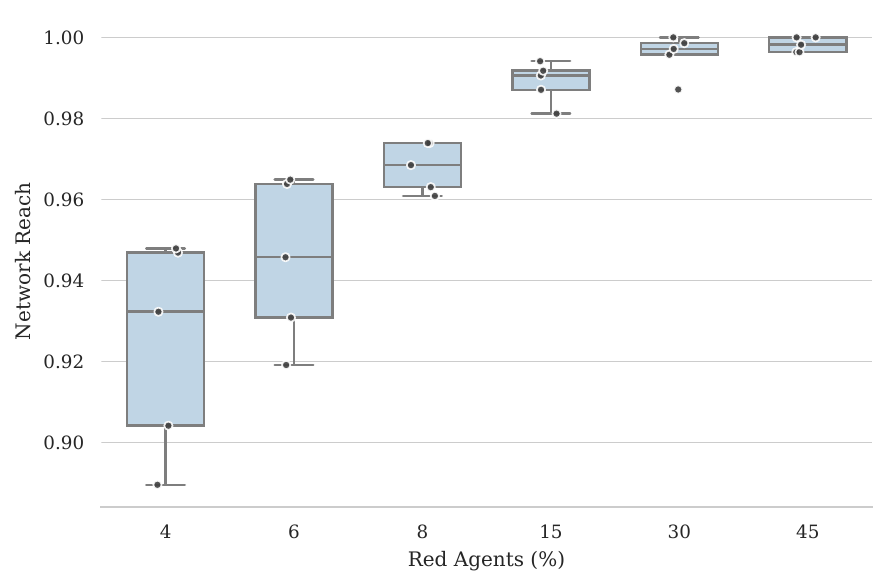}
    \caption{Counter-Narrative Reaction: Distribution of the IO reach (five simulations) per red agent configuration.}
    \label{fig:counter_reach}
\end{figure}

\textit{Counter-Narrative Reaction} is a reactive workflow in which controlled accounts reply to runtime-identified narratives with coordinated counter-messaging. Figure~\ref{fig:counter_reach} shows that this mode achieves high exposure even at low attacker presence. In four of five runs, the 4\% condition already exceeds 90\% reach, and operations with 15\% or more red-controlled accounts approach complete coverage.

This suggests that reactive attachment to already active threads is a highly efficient way to inject adversarial content into the network's attention flow. Under the tested exposure model, \textit{Counter-Narrative Reaction} is the least density-dependent workflow in terms of reach.

\begin{figure}[h!]
    \centering
    \includegraphics[width=\columnwidth]{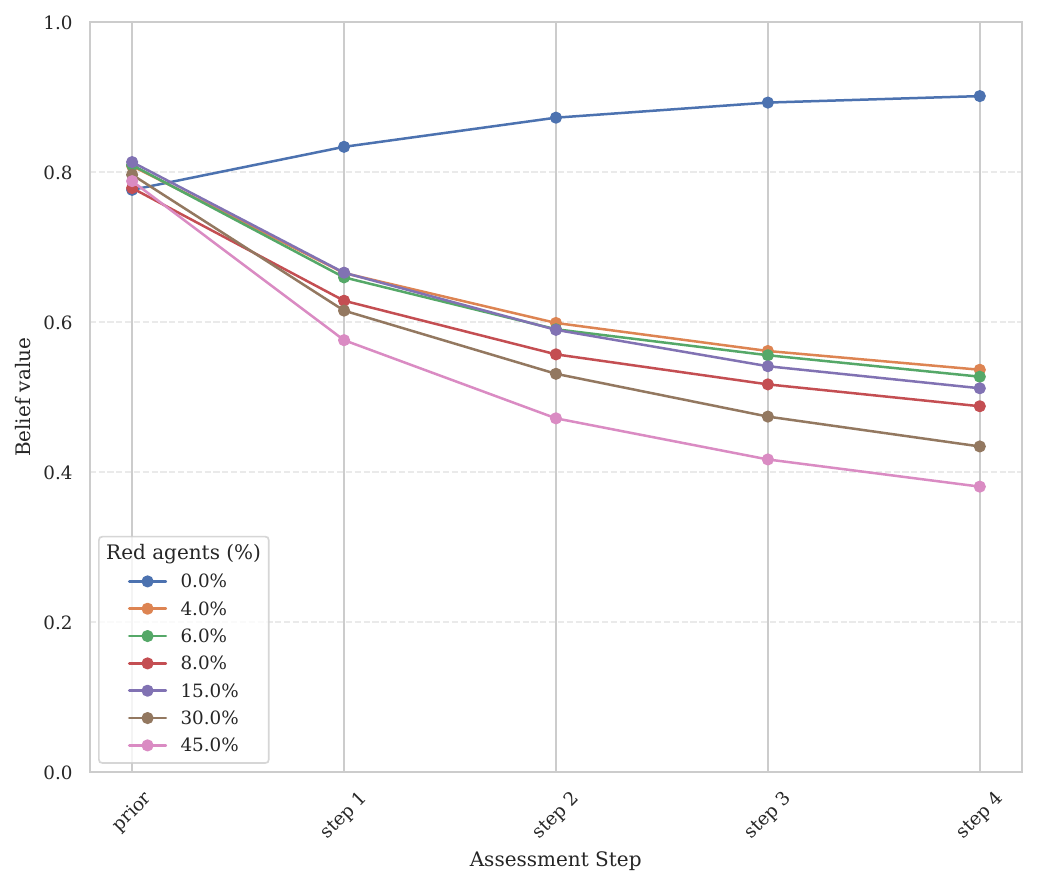}
    \caption{Counter-Narrative Reaction: Mean belief value evolution towards the original target narrative.}
    \label{fig:counter_belief}
\end{figure}

The strongest effect appears in belief change. Figure~\ref{fig:counter_belief} shows that, under the baseline, agreement with the original narrative strengthens over time. Once the workflow is activated, that trend reverses and agreement declines across all attacker densities. At high densities, the reversal is substantial. 

Within the evaluated scenario set, \textit{Counter-Narrative Reaction} yields the strongest combination of exposure and belief change. This makes it the most effective workflow in the tested conditions, but not a topic-independent or platform-independent optimum.

\subsection{Comparative view}

Taken together, the results show clear differences between the three workflows. \textit{Narrative Release} needs more attackers to become persuasive. \textit{Narrative Support} reaches many users, but changes opinions less. \textit{Counter-Narrative Reaction} performs best in both reach and opinion change by reacting to discussions that are already active.

More broadly, these results suggest that the effectiveness of AI-enabled influence operations depends not only on how many accounts an adversary controls, but also on how those accounts focus their attention. Under the simulator assumptions used here, the three workflows lead to clearly different effects.

\section{SECURITY IMPLICATIONS AND LIMITATIONS}

The results have two sides from a security perspective. On the one hand, they show that controlled simulation can be used to launch, compare, and measure AI-enabled influence operations in a repeatable way. This is valuable for security research because it enables systematic testing of offensive strategies and defensive interventions under fixed conditions. On the other hand, the same results show that these agents can achieve substantial reach and, in some cases, measurable opinion change, highlighting the potential risks of deploying similar operations on real platforms.

A first implication is that exposure and persuasion should not be treated as equivalent outcomes. Some workflows achieve broad visibility with limited additional belief change, while others combine both. For defensive evaluation, this means that reach alone is not enough: a useful assessment should also consider whether exposure leads to persuasive impact.

A second implication concerns attacker efficiency. The results suggest that workflow design matters as much as the presence of attackers. In particular, reactive strategies that attach messages to already active discussions appear especially effective. This matters for defense because it points to a practical risk: agentic campaigns may not need large attacker footprints if they can identify salient conversations and intervene at the right time.

Several limitations also apply. The experiments were conducted in synthetic online social networks rather than on live platforms, which improved control but abstracted from moderation, recommendation systems, and real-world feedback loops. The simulator also adopts a fixed follower-network exposure model, and attacker placement in the graph is not isolated, even though these factors may affect both diffusion and persuasion. In addition, the evaluated workflows do not cover the full space of influence tactics, and each workflow is tested on a different campaign-topic configuration, so workflow effects cannot be fully separated from topic sensitivity. Belief change is measured through elicited belief vectors and Jensen--Shannon divergence, which helps separate visibility from cognitive impact, but remains an elicitation-based assessment rather than a mechanistic model of cognitive updating. Finally, the threat model is better understood as AI-enabled or hybrid than as fully autonomous, since campaign goals and workflow selection remain explicitly configured while the LLM is used for bounded generation and analysis tasks.

\section{CONCLUSION}

This article presented an agent-based framework for evaluating AI-enabled influence operations in synthetic online social networks under controlled conditions. By combining a simulation environment with offensive agents, the framework enables systematic comparison of three offensive strategies through complementary measures of operation reach and audience belief change.

The results show that both workflow type and red-agent density affect reach and opinion change in synthetic audiences. \textit{Narrative Release} increases reach but needs higher densities to persuade. \textit{Narrative Support} reaches many users but produces limited additional belief change. \textit{Counter-Narrative Reaction} achieves the strongest overall effect.

Controlled simulation, therefore, provides a useful security-oriented testbed for comparing offensive strategies and supporting future evaluation of defensive interventions.

Future work should evaluate adaptive workflows, vary graph structure and attacker placement, introduce alternative exposure assumptions, and calibrate simulated operations more closely against real-world traces. It should also broaden the assessment space beyond belief elicitation to capture longer-horizon outcomes relevant to both attack analysis and defensive intervention.

\section{ACKNOWLEDGMENTS}
This work has been partially funded by the University of Murcia with the FPI/0000902983 contract.

\def\refname{REFERENCES}
\bibliographystyle{IEEEtran}
\bibliography{bilbio}

\begin{IEEEbiography}{Alejandro Buitrago L\'{o}pez}{\,} is working towards a Ph.D. in Computer Science at the University of Murcia, Spain. He obtained a B.Sc. Degree with a focus on software engineering and a M.Sc. in Big Data. He is a member of the CyberDataLab at the University of Murcia, and his research interests include data science, disinformation, and cybersecurity. Contact him at alejandro.buitragol@um.es.
\end{IEEEbiography}

\begin{IEEEbiography}{David Montoro Aguilera}{\,} is computer engineer with a focus on Computer Science. He was a member of the CyberDataLab at the University of Murcia, and his main research interests ara A.I, disinformation and machine learning. Contact him at d.montoroaguilera@um.es.
\end{IEEEbiography}

\begin{IEEEbiography}{Javier Pastor-Galindo} is Assistant Professor of Computer Science and Artificial Intelligence at the University of Murcia (Spain). His research focuses on combating online influence operations and mis/disinformation through artificial intelligence applications, cyber threat intelligence and simulation environments in both civilian and military contexts. Contact him at javierpg@um.es.
\end{IEEEbiography}

\begin{IEEEbiography}{Jos\'e~A.~Ruip\'erez-Valiente} {\,} s received his B.Eng. degree in telecommunications from Universidad Católica de San Antonio de Murcia in 2011 and a M.Eng. degree in telecommunications in 2013, together with his M.Sc. and Ph.D. degrees (2014 and 2017) in telematics from Universidad Carlos III of Madrid while conducting research with Institute IMDEA Networks in the area of applied data science.  He is currently an Associate Professor of Computer Science and Artificial Intelligence at the University of Murcia. Contact him at jruiperez@um.es.
\end{IEEEbiography}

\end{document}